\documentclass[twocolumn,english,pra,showpacs]{revtex4-1}
\usepackage[T1]{fontenc}
\usepackage[latin9]{inputenc}
\usepackage{amsmath}
\usepackage{graphicx}
\usepackage{amssymb}
\usepackage{esint}

\makeatletter
\@ifundefined{textcolor}{}
{%
 \definecolor{BLACK}{gray}{0}
 \definecolor{WHITE}{gray}{1}
 \definecolor{RED}{rgb}{1,0,0}
 \definecolor{GREEN}{rgb}{0,1,0}
 \definecolor{BLUE}{rgb}{0,0,1}
 \definecolor{CYAN}{cmyk}{1,0,0,0}
 \definecolor{MAGENTA}{cmyk}{0,1,0,0}
 \definecolor{YELLOW}{cmyk}{0,0,1,0}
 }

\makeatletter
\usepackage{dcolumn}
\usepackage{bm}
\usepackage{amsthm}\@ifundefined{definecolor}
 {\@ifundefined{definecolor}
 {\usepackage{color}}{}
}{}

\makeatother

\usepackage{babel}

\makeatother

\usepackage{babel}

\begin{document}
\newtheorem{conjecture}{Conjecture}\newtheorem{corollary}{Corollary}\newtheorem{theorem}{Theorem}
\newtheorem{lemma}{Lemma}
\newtheorem{observation}{Observation}
\newtheorem{definition}{Definition}\newtheorem{remark}{Remark}\global\long\global\long\def\ket#1{|#1 \rangle}
 \global\long\global\long\def\bra#1{\langle#1|}
 \global\long\global\long\def\proj#1{\ket{#1}\bra{#1}}

\title{Information gain versus coupling strength in quantum measurements}

\author{Xuanmin Zhu{$^{1}$$^{,}$$^{2}$$^{,}$$^{3}$}}\email{zhuxuanmin@gmail.com} \author{Yuxiang Zhang{$^{3}$$^{,}$${^4}$}} \author{Quanhui Liu$^{2}$}\email{quanhuiliu@gmail.com} \author{Shengjun Wu$^{3}$}\email{shengjun@ustc.edu.cn}

 \affiliation{$^{1}$ School of Science, Xidian University, Xi'an 710071, China \\
 $^{2}$School for Theoretical Physics, and Department of Applied Physics,
 Hunan University, Changsha 410082, China \\
 $^{3}$Hefei National Laboratory for Physical Sciences at Microscale
 and Department of Modern Physics, University of Science and Technology
 of China, Hefei, Anhui 230026, China   \\
 $^{4}$School for the Gifted Young, University of Science and Technology of China, Hefei, Anhui 230026, China}

\date{\today}

\begin{abstract}
We investigate the relationship between the information gain and the interaction strength between the quantum system and the measuring device. A strategy
is proposed to calculate the information gain of the measuring device as a function of the coupling strength. For qubit systems, we prove that the information gain increases monotonically with the coupling strength. It is obtained that the information gain of the projective measurement along the $x$ direction decreases with increasing measurement strength along the $z$ direction, and a complementarity of information gain in the measurements along those two directions is presented.
\end{abstract}
\maketitle
\section{Introduction}
In quantum information theory, quantum systems can be used to transmit classical information. But quantum systems can be neither unambiguously distinguished~\cite{holevo1,pang}, nor perfectly cloned~\cite{noclone} in general. Usually the information encoded in quantum systems cannot be transmitted without any distortion. Even if the transmitted quantum states are not disturbed during the transmission, there is a Holevo bound that limits the accessible information of the receivers~\cite{holevo,nielsen}.

The information transmission process can be described as follows: there is a classical information source which produces symbols $i= 1,...,n$ according to a probability distribution $p_1,...,p_n$. The classical information is quantified by the Shannon entropy $H(p_i)=-\sum_i p_i\mathrm{log}p_i$, where the
base of the logarithm function is 2 in this paper. The message sender Alice encodes the information into the quantum state $\rho_i$ with the probability $p_i$ where $i=1,...,n$. The receiver Bob performs a measurement described by the positive operator valued measure (POVM), $\{E_j\}=\{E_1,...,E_m\}$, to gain the information~\cite{nielsen}. If the measured state is $\rho_i$, the probability of obtaining output $j$ is $p_{j|i}=\mathrm{Tr}(E_j\rho_i)$, and $p_{ij}=p_i\mathrm{Tr}(E_j\rho_i)$. The accessible information on Bob is $I_{acc}=H(p_i)+H(p_j)-H(p_{ij})$. The Holevo bound is $\chi=S(\rho)-\sum_i p_i S(\rho_i)$~\cite{holevo,nielsen}, we have $I_{acc} \leq S(\rho)-\sum_i p_i S(\rho_i)$, where $\rho=\sum_i p_i \rho_i$, and $S(\rho)$ is the von Neumann entropy of the state $\rho$. From the properties of the von Neumann entropy~\cite{nielsen,wehrl}, we get that $I_{acc}\leq S(\rho)-\sum_i p_i S(\rho_i)\leq H(p_i)$ which means that the accessible information on the receiver is less than the original information.

From another perspective, the Holevo bound $\chi$ is equal to the mutual information $S(A:B)$ of the bipartite state $\rho_{AB}=\sum_i p_i |i\rangle_{A}\langle i| \otimes \rho_{iB}$, where $\{|i\rangle_{A}\}$ is an orthonormal basis, and $\rho_{iB}=\rho_i$ is the transmitted state. From the aspect of the quantum correlation theory~\cite{discord1, discord2, luo, luo1, wu}, the mutual information $S(A:B)$ is considered to be the total correlation of subsystems $A$ and $B$. The maximal classical information that Bob can gain from states $\{\rho_{iB}\}$ is the classical correlation $I_{\max}(A:B)$ of state $\rho_{A,B}$ which is defined as $I_{\max}(A:B)=\max_{\{E_j\}}\{H(p_i)+H(p_j)-H(p_{ij})\}$. It has been proven that $I_{\max}(A:B) \leq S(A:B)$~\cite{wu,barnett}, so we have $I_{\max}(A:B) \leq \chi$, which provides an alternative proof of the Holevo bound.

In actual experiments, for the technical limits or some special purposes, the interactions between quantum systems and measuring devices may not be very strong. So, it is interesting to study the relationship between the interaction strength and the information gain of the measuring device. In this paper, we calculate the value of the information gain as a function of the coupling strength. From our intuition, the information gain of a measuring device increases with the coupling strength between the device and the quantum system. For qubit systems, we prove that our intuition is actually true. We also prove that the information gain of the projective measurement along the $x$ direction decreases with an increase in the measurement strength along the $z$ direction. Based on the monotonicity, we obtain a complementarity of the information gain in the measurements along two perpendicular directions.
\section{The information gain}
The quantum states sent by Alice constitute an ensemble $\{p_i, \rho_{iB}\}$, specified by Alice sending state $\rho_i$ with probability $p_i$, where
$i = 1, ..., n$. The ensemble can be described by the density matrix $\rho_{B}=\sum_{i}p_i \rho_{iB}$. The state of the measuring device is $|\Phi\rangle_{D}$; the interaction between the quantum systems and the measuring device is assumed to be impulsive which can be described as~\cite{w1,w2,w3,w4}
\begin{equation}\label{eq01}
 H_{int}=g\delta(t-t_0)B\otimes D,
\end{equation}
where $B$ is an observable operator of the quantum systems, $D$ is an operator of the measuring device, and $g$ is the coupling strength with the assumption that $g\geq 0$. We introduce a fictitious auxiliary system $A$ which can be thought of as the "preparation" system. The auxiliary system has an orthonomal basis $\{|i\rangle_{A}\}$ whose elements correspond to the labels $1,2,...,n$ on the possible preparations for the transmitted system, $B$. The states of $A$ can be considered as the memory of the original information source. Before the interaction, the overall state of $A$, $B$, and the measuring device $D$ is
\begin{equation}\label{eq02}
\rho_{ABD}=\sum_i p_i |i\rangle_{A}\langle i|\otimes \rho_{iB}\otimes |\Phi\rangle_D\langle\Phi|.
\end{equation}
After the interaction the overall state evolves into
\begin{equation}\label{eq03}
\rho_{ABD}'=\sum_i p_i |i\rangle_{A}\langle i|\otimes U \rho_{iB} \otimes |\Phi\rangle_D\langle\Phi|U^{\dagger},
\end{equation}
where $U=e^{-i\int H_{int} \mathrm{d}t}=e^{-igB\otimes D}$, with $\hbar=1$ throughout this paper. It is assumed that the complete orthonormal eigenstates of the observable $B$ are $\{|b_m\rangle\}$, and the corresponding eigenvalues are $\{b_m\}$. States $\rho_{iB}$ and $\rho_{B}$ can be written as
\begin{equation}\label{eq04}\begin{split}
\rho_{iB}=\sum_{mn}\rho_{mn}^{i}|b_m\rangle\langle b_n|,\\
\rho_{B}=\sum_{mn}\rho_{mn}|b_m\rangle\langle b_n|.
\end{split}\end{equation}
After the interaction, state $\rho_{iBD}=\rho_{iB}\otimes|\Phi\rangle_D\langle\Phi|$ evolves into
\begin{equation}\label{eq05}
\rho_{iBD}'=\sum_{mn}\rho_{mn}^{i}e^{-igb_m D}|b_m\rangle\langle b_n|\otimes|\Phi\rangle_D\langle\Phi|e^{igb_n D}.
\end{equation}
We get the measuring device's state
\begin{equation}\label{eq06}
\rho_{iD}'=\mathrm{tr}_{B}(\rho_{iBD}')=\sum_{m}\rho_{mm}^{i}e^{-igb_m D}|\Phi\rangle_D\langle\Phi|e^{igb_m D}.
\end{equation}
Similarly, we can obtain the final overall state of the measuring device and the information source,
\begin{equation}\label{eq07}
\rho_{AD}'=\mathrm{tr}_{B}(\rho_{ABD}')=\sum_i p_i |i\rangle_{A}\langle i|\otimes\rho_{iD}'.
\end{equation}
The mutual information $S(A:D)$ of the measuring device and system $A$ represents the correlation of the measuring device and the information source~\cite{luo2}, so we define the information gain of the measuring device,
\begin{equation}\label{eq08}\begin{split}
I_a=S(A:D)&=S(\rho_A')+S(\rho_D')-S(\rho_{AD}')\\
&=S(\rho_D')-\sum_i p_i S(\rho_{iD}'),
\end{split}\end{equation}
where $\rho_A'=\rho_A=\sum_i p_i |i\rangle_{A}\langle i|$ is the total density matrix of the information source, and $\rho_D'=\sum_i p_i \rho_{iD}'$ is the total density matrix of the measuring device. From another perspective, the information gain $I_a$ is the Holevo bound for the case that the classical information is encoded in the measuring device's states $\{\rho_{iD}'\}$ with the probabilities $\{p_i\}$.

Now, we prove that the information gain $I_a$ is less than the Holevo bound $\chi=S(\rho_B)-\sum_i p_i S(\rho_{iB})$. From the theory of relative entropy~\cite{nielsen,vedral}, we have
\begin{equation}\label{eq09}\begin{split}
\chi &=S(\rho_{AB}\otimes |\Phi\rangle_D\langle\Phi| \parallel \rho_{A}\otimes \rho_{B}\otimes|\Phi\rangle_D\langle\Phi|)\\
&=S(\rho_{ABD}' \parallel U_{ABD} \rho_{A}\otimes \rho_{B}\otimes|\Phi\rangle_D\langle\Phi|U_{ABD}^{\dagger})
\end{split}\end{equation}
where $U_{ABD}=I_{A}\otimes e^{-i\int H_{int} \mathrm{d}t}$ is a unitary operator and $\rho_{ABD}'=U_{ABD}\rho_{AB}\otimes |\Phi\rangle_D\langle\Phi |U_{ABD}^{\dagger}$. Based on the monotonicity of relative entropy~\cite{nielsen,vedral}, we obtain
\begin{equation}\label{eq10}\begin{split}
\chi&\geq S(\mathrm{tr}_{B}(\rho_{ABD}') \parallel\mathrm{tr}_{B}( U_{ABD}\rho_{A}\otimes \rho_{B}\otimes|\Phi\rangle_D\langle\Phi| U^{\dagger}_{ABD}))\\
&=S(\rho_{AD}' \parallel \rho_{A}'\otimes \rho_{D}') \\
&=S(\rho_D')-\sum_i p_i S(\rho_{iD}')=I_a.
\end{split}\end{equation}
Thus we obtain that the information gain $I_{a}$ is less than the Holevo bound $\chi$.

Without loss of generality, the initial state of the measuring device is assumed to be a Gaussian wave function centered on $q=0$
\begin{equation}\label{eq11}
\Phi(q)=\frac{1}{ (2 \pi\Delta^{2})^{\frac{1}{4}}}\exp({-\frac{q^2}{4\Delta^2}}),
\end{equation}
where the standard deviation $\Delta q= \Delta$. The original density matrix of the measuring device is
\begin{equation}\label{eq12}
\rho_{D}=\frac{1}{ (2 \pi\Delta^{2})^{\frac{1}{2}}}\int\int e^{-\frac{q^2}{4\Delta^2}}e^{-\frac{q'^2}{4\Delta^2}}|q\rangle\langle q'|\mathrm{d}q\mathrm{d}q'.
\end{equation}
The interaction Hamiltonian considered is $H_{int}=g\delta(t-t_0)B\otimes p$.  From Eq. (\ref{eq06}), we obtain the density matrix $\rho_{iD}'$ is
\begin{equation}\label{eq13}\begin{split}
\rho_{iD}'&=\sum_{m}\rho_{mm}^{i}e^{-igb_m p}\rho_{D}e^{igb_m p}\\
&=\sum_{m}\frac{\rho_{mm}^{i}}{ (2 \pi\Delta^{2})^{\frac{1}{2}}}\int\int e^{-\frac{(q-gb_m)^2}{4\Delta^2}}e^{-\frac{(q'-gb_m)^2}{4\Delta^2}} |q\rangle\langle q'| \mathrm{d}q\mathrm{d}q'.
\end{split}\end{equation}
Since the $\rho_{iD}'$ is a continuum variable density matrix, it is not easy to calculate its von Neumann entropy directly. We can introduce an auxiliary system $R$ to purify the state of the measuring device, and the state of the combined system is
\begin{equation}\label{eq14}
|\Psi_i\rangle_{DR}=\sum_m\frac{\sqrt{\rho_{mm}^{i}}}{ (2 \pi\Delta^{2})^{\frac{1}{4}}}\int e^{-\frac{(q-gb_m)^2}{4\Delta^2}}|m\rangle_R |q\rangle \mathrm{d}q,
\end{equation}
where $\{|m\rangle_{R}\}$ is an orthonormal basis of the auxiliary system, and $\rho_{iD}'=\mathrm{tr}_R(|\Psi_i\rangle_{DR}\langle\Psi_i|)$. As $|\Psi_i\rangle_{DR}$ is a pure state, we have
\begin{equation}\label{eq15}
S(\rho_{iD}')=S(\rho_{iR}),
\end{equation}
and the density matrix $\rho_{iR}$ is
\begin{equation}\label{eq16}\begin{split}
\rho_{iR}=\mathrm{tr}_{D}(|\Psi_i\rangle_{DR}\langle\Psi_i|)=\sum_{mn} \rho_{mn}^{iR}|m\rangle\langle n|.
\end{split}\end{equation}
We can obtain the matrix elements of $\rho_{iR}$
\begin{equation}\label{eq17}\begin{split}
\rho_{mn}^{iR}&=\frac{\sqrt{\rho_{mm}^{i}\rho_{nn}^{i}} }{ (2 \pi\Delta^{2})^{\frac{1}{2}}}\int\int e^{-\frac{(q-gb_m)^2}{4\Delta^2}}e^{-\frac{(q'-gb_n)^2}{4\Delta^2}}\langle q'| q\rangle\mathrm{d}q\mathrm{d}q'\\
&=\sqrt{\rho_{mm}^{i}\rho_{nn}^{i}} e^{-\frac{g^2(b_m-b_n)^2}{8\Delta^2}}.
\end{split}\end{equation}
It can be seen that the dimension of the matrix $\rho_{iR}$ is the same with the observable $B$. By a similar derivation, we obtain
\begin{equation}\label{eq18}
S(\rho_D')=S(\rho_{R})=S(\sum_{mn} \rho_{mn}^{R}|m\rangle\langle n|),
\end{equation}
and the matrix element $\rho_{mn}^{R}=\sqrt{\rho_{mm}\rho_{nn}} e^{-\frac{g^2(b_m-b_n)^2}{8\Delta^2}}$. So we can get the von Neumann entropy of the measuring device by calculating the entropy of the auxiliary system $R$. From Eqs. (\ref{eq08}), (\ref{eq15}), and (\ref{eq18}), the information gain of the measuring device is
\begin{equation}\label{eq19}
I_a=S(\rho_{R})-\sum_{i}p_iS(\rho_{iR}).
\end{equation}

When the coupling strength is strong (i.e., $g\gg \Delta$), and the eigenvalues of $B$ are nondegenerate, we will prove that the information gain is equal to the information $I_p$ extracted by the projective measurement along the orthonormal eigenstates $\{|b_{m}\rangle \}$ of $B$. The information $I_p$
obtained in the projective measurement along the basis $\{|b_{m}\rangle]\}$ is
\begin{equation}\label{new1}
I_p=H(p_m)+H(p_i)-H(p_{im}).
\end{equation}
where $H(p_m)=-\sum_m p_m\mathrm{log}p_m$ is the Shannon entropy, $p_m=\mathrm{tr}(|b_m\rangle\langle b_m|\rho_{B})$, and the joint probability $p_{im}=p_i\mathrm{tr}(|b_m\rangle\langle b_m|\rho_{iB})$.

When $g\gg \Delta$, and $m\neq n$, we have $e^{-{g^2(b_m-b_n)^2}/{8\Delta^2}} \to 0 $, the nondiagonal elements of matrices $\rho_{iR}$ and $\rho_{R}$ are approximatively equal to 0, and we have
\begin{equation}\label{eq20}
\rho_{iR}\approx \sum_{m} \rho_{mm}^{i}|m\rangle\langle m|, \rho_{R}\approx \sum_{m} \rho_{mm}|m\rangle\langle m|.
\end{equation}
From Eq. (\ref{eq04}), we have $\rho_{iR}=\rho_{iB}'=\sum_m \langle b_m|\rho_{iB}|b_m\rangle |b_m\rangle\langle b_m|$, and $\rho_{R}=\rho_{B}'=\sum_m \langle b_m|\rho_{B}|b_m\rangle |b_m\rangle\langle b_m|$ which are the states after the projective measurements on states $\rho_{iB}$ and $\rho_{B}$ along the basis $\{|b_m\rangle\}$, respectively. From Eq. (\ref{eq19}), the information gain is
\begin{equation}\label{eq21}\begin{split}
I_a&=S(\rho_{B}')-\sum_i p_i S(\rho_{iB}')\\
&=H(p_m)-\sum_i p_i H(p_{im|i})\\
&=H(p_m)+H(p_i)-H(p_{im})=I_p.
\end{split}\end{equation}
Thus we have proved that the information gain $I_a$ equals the information obtained by measuring the states transmitted $\{\rho_{iB}\}$ along the basis $\{|b_m\rangle\}$. It can be seen that when the coupling strength is large, the information gain of the measuring device is equal to the information obtained in the ideal projective measurement, which is consistent with our expectation.
\section{The monotonicity of the information gain}
Now we study the monotonicity of the information gain $I_a$ and the coupling strength $g$ when the transmitted quantum systems are qubits. For two-dimensional systems, the orthonomal eigenstates of observable $B$ can be denoted as $\{|0\rangle, |1\rangle\}$, and without loss of generality, the corresponding eigenvalues are assumed to be $\{1,-1\}$. The general state of a qubit can be represented as a point in the Bloch sphere~\cite{nielsen}. We can use three parameters, $r$ (radius), $\theta$ (polar angle), and $\phi$ (phase angle), to define a qubit state, where $0 \leq r \leq 1$, $0\leq \theta < \pi$, and $0 \leq \phi< 2\pi$. In the representation $\{|0\rangle, |1\rangle\}$, the transmitted state $\rho_{iB}$ can be written as
\begin{equation}\label{eq22}
\rho_{iB}=\left(
  \begin{array}{ccc}
    \frac{1+r_i\cos{\theta_i}}{2} & \frac{r_i\sin{\theta_i}e^{-i\phi_i}}{2} \\
    \frac{r_i\sin{\theta_i}e^{i\phi_i}}{2} & \frac{1-r_i\cos{\theta_i}}{2} \\
  \end{array}
\right).
\end{equation}
Then $\rho_{11}^{i}= \frac{1+r_i\cos{\theta_i}}{2}$ and $\rho_{22}^{i}= \frac{1-r_i\cos{\theta_i}}{2}$, and from Eqs. (\ref{eq16}) and (\ref{eq17}), we have
\begin{equation}\label{eq23}
 \rho_{iR}=\left(
  \begin{array}{ccc}
    \frac{1+r_i\cos{\theta_i}}{2} & \sqrt{\frac{1-r_i^2\cos^2{\theta_i}}{4}}e^{-\frac{g^2}{2\Delta^2}} \\
   \sqrt{\frac{1-r_i^2\cos^2{\theta_i}}{4}}e^{-\frac{g^2}{2\Delta^2}} & \frac{1-r_i\cos{\theta_i}}{2} \\
  \end{array}
\right).
\end{equation}
Then we obtain the entropy of the state $\rho_{iR}$ as
\begin{equation}\label{eq24}
S(\rho_{iR})=H_B\left(\lambda_i\right),
\end{equation}
where $\lambda_i=({1+({r_i^2\cos^2{\theta_i}+(1-r_i^2\cos^2{\theta_i})e^{-\frac{g^2}{\Delta^2}}})^{1/2}})/{2}$, and $H_B(\lambda_i)=-\lambda_i\mathrm{log}\lambda_i-(1-\lambda_i)\mathrm{log}(1-\lambda_i)$ is the binary Shannon entropy. By similar calculations, we have
\begin{equation}\label{eq25}
S(\rho_{R})=H_B\left(\frac{1+s^{1/2}}{2}\right),
\end{equation}
where $s=(\sum_i p_i r_i\cos{\theta_i})^2+(1-(\sum_i p_i r_i\cos{\theta_i})^2)e^{-\frac{g^2}{\Delta^2}}$. From Eqs. (\ref{eq19}), (\ref{eq24}), and (\ref{eq25}), the information gain of the measuring device is
\begin{equation}\label{eq26}
I_{a}=H_B\left(\frac{1+s^{1/2}}{2}\right)-\sum_i p_i H_B\left(\lambda_i\right).
\end{equation}

In the following theorem, we present that the information gain $I_a$ increases with the coupling strength.
\begin{theorem} \label{thm01}
When the transmitted systems are qubits, the information gain $I_a$ monotonically increases with the coupling strength $g$.
\end{theorem}
The proof of this theorem is given in the Appendix.

Now we consider the case when information eavesdroppers are in. In this case, an eavesdropper named Eve intercepts the qubits which are transmitted from Alice to Bob, performs a measurement on the qubits for extracting the information sent to Bob, and resends the states to Bob. The interaction Hamiltonian between the quantum systems and Eve's measuring device is
\begin{equation}\label{eq27}
H_{int}=g\delta(t-t_0)\sigma_z\otimes p.
\end{equation}
Without loss of generality, we have assumed that this measurement is along the $z$ direction. The information gain $I_{a,z}$ of Eve is given by Eq. (\ref{eq26}). After the measurement performed by Eve, the state $\rho_{iB}$ given in Eq. (\ref{eq22}) is changed into
\begin{equation}\label{eq28}\begin{split}
\rho_{iB}'&=\mathrm{tr}_D(\rho_{iBD}')\\
&=\left(
  \begin{array}{ccc}
    \frac{1+r_i\cos{\theta_i}}{2} &\frac{ r_i\sin{\theta_i}e^{-i\phi_i}e^{-\frac{g^2}{2\Delta^2}}}{2} \\
    \frac{r_i\sin{\theta_i}e^{i\phi_i}e^{-\frac{g^2}{2\Delta^2}}}{2} & \frac{1-r_i\cos{\theta_i}}{2} \\
  \end{array}
\right),
\end{split}\end{equation}
and the total density matrix of the ensemble evolves into
\begin{equation}\label{eq29}\begin{split}
\rho_{B}'&=\sum_i p_i\rho_{iB}\\
&=\left(
  \begin{array}{ccc}
    \frac{1+\sum_i p_i r_i\cos{\theta_i}}{2} & \frac{\sum_i p_ir_i\sin{\theta_i}e^{-i\phi_i}e^{-\frac{g^2}{2\Delta^2}}}{2} \\
    \frac{\sum_i p_i r_i\sin{\theta_i}e^{i\phi_i}e^{-\frac{g^2}{2\Delta^2}}}{2} & \frac{1-\sum_i p_i r_i\cos{\theta_i}}{2} \\
  \end{array}
\right).
\end{split}\end{equation}
Finally, the legitimate receiver Bob performs a projective measurement on his received quantum system along the $x$ direction to gain the information from Alice. The projective measurement operators are $\{|+\rangle\langle+|, |-\rangle\langle-|\}$, where $|+\rangle=\frac{1}{\sqrt{2}}(|0\rangle+|1\rangle)$ and $|-\rangle=\frac{1}{\sqrt{2}}(|0\rangle-|1\rangle)$. The information gain $I_{a,x}$ of Bob is
\begin{equation}\label{eq30}
I_{a,x}=H(p_i)+H(p_j)-H(p_{ij}),
\end{equation}
where $j=+,-$, $p_+=\mathrm{tr}(|+\rangle\langle+|\rho_{B}')$, $p_-=\mathrm{tr}(|-\rangle\langle-|\rho_{B}')$, $p_{i+}=p_i\mathrm{tr}(|+\rangle\langle+|\rho_{iB}')$, and $p_{i-}=p_i\mathrm{tr}(|-\rangle\langle-|\rho_{iB}')$.
From Eqs. (\ref{eq28}), (\ref{eq29}), and (\ref{eq30}), we have
\begin{equation}\label{eq31}\begin{split}
I_{a,x}=&H_B\left(  \frac{1+\sum_i p_i r_i \sin{\theta_i}\cos{\phi_i}e^{-\frac{g^2}{2\Delta^2}}}{2}   \right)\\
&-\sum_i p_i H_B\left(  \frac{1+ r_i \sin{\theta_i}\cos{\phi_i}e^{-\frac{g^2}{2\Delta^2}}}{2}   \right).
\end{split}\end{equation}
Now, we give a theorem to show that the information gain $I_{a,x}$ decreases with an increase in the coupling strength $g$.
\begin{theorem} \label{thm02}
For qubit systems, the information gain $I_{a,x}$ of the projective measurement along the $x$ direction monotonically decreases with the measurement coupling strength $g$ along the $z$ direction.
\end{theorem}
 This monotonicity is consistent with the widely studied information-disturbance trade-off relation~\cite{fuchs, maccone, barbieri, luo3}. The proof of this theorem is given in the Appendix.
\section{Complementarity of the information gain}
From Theorem 1, we know that the information gain $I_{a,z}$ of Eve increases with $g$. For $g\to +\infty$, from Eq. (\ref{eq21}), the information gain is equal to the information gain of the projective measurement along the basis $\{|0\rangle, |1\rangle\}$, which is
\begin{equation}\label{eq32}\begin{split}
I_{z}&=H(p_i)+H(p_{j,z})-H(p_{ij,z})\\
&=H(p_{j,z})-\sum_i p_i H(p_{j,z|i}),
\end{split}\end{equation}
where $p_{1,z}=\mathrm{tr}(|0\rangle\langle0|\rho_B )$, $p_{2,z}=\mathrm{tr}(|1\rangle\langle1|\rho_B )$, $p_{i1,z}=p_i\mathrm{tr}(|0\rangle\langle0|\rho_{iB})$, $p_{i2,z}=p_i\mathrm{tr}(|1\rangle\langle1|\rho_{iB})$, and we have $I_{a,z}\leq I_{z}$. In Theorem 2, it is shown that the information gain $I_{a,x}$ decreases with the coupling strength $g$, when $g=0$, we have
\begin{equation}\label{eq33}\begin{split}
I_{x}&=H(p_i)+H(p_{j,x})-H(p_{ij,x})\\
&=H(p_{j,x})-\sum_i p_i H(p_{j,x|i}),
\end{split}\end{equation}
where $p_{1,x}=\mathrm{tr}(|+\rangle\langle+|\rho_B )$, $p_{2,x}=\mathrm{tr}(|-\rangle\langle-|\rho_B )$, $p_{i1,x}=p_i\mathrm{tr}(|+\rangle\langle+|\rho_{iB})$, $p_{i2,x}=p_i\mathrm{tr}(|-\rangle\langle-|\rho_{iB})$, and we have $I_{a,x}\leq I_x$. Then we obtain
\begin{equation}\label{eq34}\begin{split}
I_{a,z}+I_{a,x} & \leq I_{z} +I_{x}\\
&=H(p_{j,z})+H(p_{j,x})-\sum_i p_i(H(p_{j,z|i})+H(p_{j,x|i})),
\end{split}\end{equation}
where the conditional probabilities $p_{1,z|i}=\mathrm{tr}(|0\rangle\langle0|\rho_{iB})$, $p_{2,z|i}=\mathrm{tr}(|1\rangle\langle1|\rho_{iB})$, $p_{1,x|i}=\mathrm{tr}(|+\rangle\langle+|\rho_{iB})$, and $p_{2,x|i}=\mathrm{tr}(|-\rangle\langle-|\rho_{iB})$. From the entropic uncertainty
relation given in~\cite{berta} and \cite{guo}, we have
\begin{equation}\label{eq35}
H(p_{j,z|i})+H(p_{j,x|i})\geq 1+S(\rho_{iB}).
\end{equation}
As $\{p_{j,z}\}$ and $\{p_{j,x}\}$ are the two-outcome probability distributions, we have $H(p_{j,z})+H(p_{j,x})\leq 2$, and from Eqs. (\ref{eq34}) and (\ref{eq35}), we obtain
\begin{equation}\label{eq36}
I_{a,z}+I_{a,x} \leq 1- \sum_i p_i S(\rho_{iB}),
\end{equation}
which is a complementarity of the information gain of the measurements along two mutually unbiased bases~\cite{ivonovic,wu1,wu2}. By complementarity, we mean that the more information eavesdropper Eve extracted from the measurement along the $z$ direction, the less information Bob could gain by the projective measurement along the $x$ direction. Numerical calculations indicate that there is the bound $I_{a,z}+I_{a,x}\leq \chi =S(\rho_B)-\sum_i p_i S(\rho_{iB})$, which is much tighter than the one given in Eq. (\ref{eq36}). Unfortunately, we do not know how to prove this inequality.

Now, we give a simple example to show the complementarity of $I_{a,z}$ and $I_{a,x}$. In the BB84 quantum key distribution protocol~\cite{bb84}, Alice sends the states $\{|0\rangle, |1\rangle, |+\rangle, |-\rangle\}$ with equal probability, and the Holevo bound is $\chi=1$. By simple calculation, we obtain the information gain,
\begin{equation}\label{eq37}\begin{split}
I_{a,z}=\frac{1}{2}H_B \left( \frac{1+e^{-g^2/2\Delta^2}}{2} \right), \\
I_{a,x}=\frac{1}{2}-\frac{1}{2}H_B \left( \frac{1+e^{-g^2/2\Delta^2}}{2} \right),
\end{split}\end{equation}
and $I_{a,z}+I_{a,x}=\frac{1}{2} < \chi$. In Fig. 1, the relationship between the information gain and the coupling strength $g$ is depicted. We can see that the information gain $I_{a,z}$ increases with $g$, while $I_{a,x}$ decreases with the value of $g$. This means that the measurement performed by Eve along the $z$ direction destroys the information that Bob could gain in the projective measurement along the $x$ direction.
\begin{figure}[t]
 \centering \includegraphics[scale=0.5]{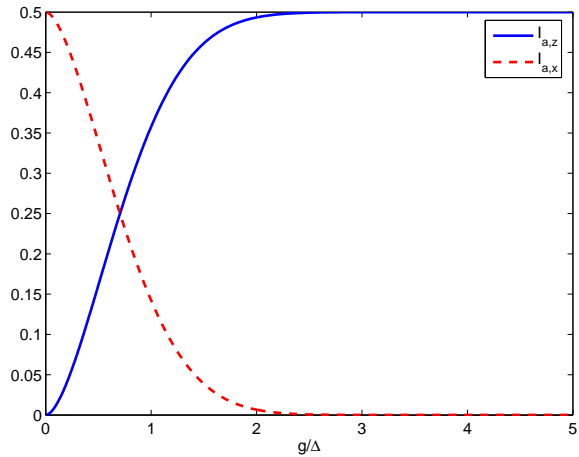} \caption{(Color online)Information gain $I_{a,z}(g)$ and $I_{a,x}$ for the BB84 protocol. }
\label{fig:01}
\end{figure}
\section{conclusions}
In conclusion, we have studied the relationship between information gain and measurement coupling strength. For a finite interaction, the information gain of the measuring device is calculated when the measuring device's states are of the Gaussian type. When the coupling strength is high, we have shown that the information gain of the measuring device is equal to the information obtained in the projective measurement. It has been proved that the information gain increases with the coupling strength $g$ monotonously for qubit systems. Complementarity of the information obtained in the measurements along two different mutually unbiased bases is given. The research in this paper is useful for evaluating the information gain in finite-interaction measurements.

\section*{Acknowledgments}
This work was financially supported by the National Natural Science Foundation of China (Grants No. 11075148, and No. 11175063).

\section{Appendix}
\subsection{Proof of Theorem 1}

\begin{proof}
Here, we show that the information gain $I_a$ is a monotonic function of $g$. As $g\geq 0$, let $t=\frac{g^2}{\Delta^2}$, $s=(\sum_i p_i r_i\cos{\theta_i})^2+(1-(\sum_i p_i r_i\cos{\theta_i})^2)e^{-t}$, and $s_i=(r_i\cos{\theta_i})^2+(1-( r_i\cos{\theta_i})^2)e^{-t}$, we have
\begin{equation}\label{a01}\begin{split}
F&=\frac{\mathrm{d}I_a}{\mathrm{d}t} \\
&=\frac{(1-(\sum_i p_i r_i\cos{\theta_i})^2)e^{-t}}{4s^{1/2}}\mathrm{log}\frac{1+s^{1/2}}{1-s^{1/2}}\\
&-\sum_i p_i \frac{(1-(r_i\cos{\theta_i})^2)e^{-t}}{4s_i^{1/2}}\mathrm{log}\frac{1+s_i^{1/2}}{1-s_i^{1/2}}
\end{split}\end{equation}
To show the monotonicity of $I_a$, we only need to prove that $F\geq 0$. Let $a=e^{-t}$, and we define a function
\begin{equation}\label{a02}
h(x)=\frac{(1-x^2)a}{4(x^2+(1-x^2)a)^{1/2}}\mathrm{log}\frac{1+(x^2+(1-x^2)a)^{1/2}}{1-(x^2+(1-x^2)a)^{1/2}},
\end{equation}
where $x\in[-1,1]$, we get
\begin{equation}\label{a03}
F=h(\sum_i p_i r_i\cos{\theta_i})-\sum_i p_i h(r_i \cos{\theta_i}).
\end{equation}
Since $\sum_i p_i =1$, if we could prove that $h(x)$ is a concave function, we will get $F\geq0$. The second derivative of $h(x)$ is
\begin{equation}\label{a04}\begin{split}
\frac{\mathrm{d}^2h(x)}{\mathrm{d}x^2}= \frac{(1-a)^2aC(x,a)}{4(x^2-1)(a+x^2-ax^2)^{1/2}D(a,x)\ln 2},
\end{split}\end{equation}
where $C(x,a)=2(a+x^2-ax^2)^{1/2}(2x^2+a(x^2-1))+(x^2-1)(2x^2+a^2(x^2-1)-a(1+3x^2))\ln \frac{1+(x^2+(1-x^2)a)^{1/2}}{1-(x^2+(1-x^2)a)^{1/2}}$ and $ D(a,x)=(a + x^2 - 2 a x^2 + a^2 (-1 + x^2))^2$; Let $w=(a+x^2-ax^2)^{1/2}$, we have $0\leq w\leq 1$ and $x^2=\frac{w^2-a}{1-a}$. Let
\begin{equation}\label{a05}\begin{split}
G(w,a)&=\frac{C(x,a)}{x^2-1}\\
&=\frac{2(2+a)w^3-6aw}{w^2-1}+((2-a)w^2-3a)\ln\frac{1+w}{1-w}.
\end{split}\end{equation}
For a fixed value of $w$, we search the extreme value of $G(w,a)$, from
\begin{equation}\label{a06}
\frac{\partial G(w,a)}{\partial a}=\frac{2w^3-6w}{w^2-1}-(3+w^2)\ln \frac{1+w}{1-w}=0,
\end{equation}
we obtain
\begin{equation}\label{a07}
\mathrm\ln\frac{1+w}{1-w}=\frac{2w^3-6w}{(w^2-1)(3+w^2)}.
\end{equation}
Substituting this solution into Eq. (\ref{a05}), the extreme value of this function is
\begin{equation}\label{a08}
G(w,a)_{\mathrm{ext}}=\frac{8w^5}{(w^2-1)(3+w^2)}.
\end{equation}
Since $0\leq w\leq 1$, we have $G(w,a)_{\mathrm{ext}} \leq 0$. As $0<a=e^{-t}\leq 1$, for $a=0$, we have
\begin{equation}\label{a09}
G(w,0)=\frac{2w^2}{w^2-1}(2w-(1-w^2)\ln\frac{1+w}{1-w} ).
\end{equation}
Let $K(w)=2w-(1-w^2)\ln\frac{1+w}{1-w}$, the first derivative of $K$ is
\begin{equation}\label{a10}
\frac{\mathrm{d}K}{\mathrm{d}w}=2w\ln \frac{1+w}{1-w}.
\end{equation}
Since $0\leq w\leq 1$, we have $\frac{\mathrm{d}K}{\mathrm{d}w}\geq 0$, as $K(0)=0$, so $K(w)\geq 0$. From Eq. (\ref{a09}), we have $G(w,0)\leq 0$. When $a=1$, we obtain
\begin{equation}\label{a11}
G(w,1)=6w+(w^2-3)\ln \frac{1+w}{1-w}.
\end{equation}
The partial derivative of $G(w,1)$ is
\begin{equation}\label{a12}\begin{split}
\frac{\partial G(w,1)}{\partial w} &=\frac{2w}{w^2-1}(2 w-(1-w^2)\ln \frac{1+w}{1-w})\\
&=\frac{2w}{w^2-1}K.
\end{split}\end{equation}
Since $K \geq 0$ and $0\leq w\leq 1$, so $\frac{\partial G(w,1)}{\partial w}\leq 0$, and as $G(0,1)=0$, we have $G(w,1)\leq 0$. Now we have proved that $G(w,0)\leq 0$, $G(w,1)\leq 0$, and the extreme value $G(w,a)_{\mathrm{ext}}\leq 0$, and as the function $G(w,a)$ is a continuum function of $a$, $0\leq w\leq 1$, and $0 <a \leq 1$, we have $G(w,a) \leq 0$. From Eqs. (\ref{a04}) and (\ref{a05}), we have
\begin{equation}\label{a13}
\frac{\mathrm{d}^2h(x)}{\mathrm{d}x^2}\leq 0,
\end{equation}
and $h(x)$ is a concave function. From Eq. (\ref{a03}), we have $F \leq 0$. Thus we have proved Theorem 1.
\end{proof}

\subsection{Proof of Theorem 2}

\begin{proof}
 As $g\geq 0$, let $t=\frac{g^2}{\Delta^2}$, $s=\sum_i p_i r_i \sin{\theta_i}\cos{\phi_i}$, and $s_i=r_i \sin{\theta_i}\cos{\phi_i}$, we have
\begin{equation}\label{a14}\begin{split}
W=\frac{\mathrm{d}{I_{a,x}}}{\mathrm{d}t}=&\frac{s e^{-t/2}}{4}\mathrm{log}\frac{1+s e^{-t/2}}{1-s e^{-t/2}}\\
&-\sum_i p_i\frac{ s_i e^{-t/2}}{4}\mathrm{log}\frac{1+s_i e^{-t/2}}{1- s_i e^{-t/2}}.
\end{split}\end{equation}
To show the monotonicity of $I_{a,x}$, we only need to prove that $W \leq 0$. We define a function $f(x)=\frac{xe^{-t/2}}{4}\mathrm{log}\frac{1+xe^{-t/2}}{1-xe^{-t/2}}$, where $x\in[-1,1]$. We have
\begin{equation}\label{a15}
W=f(\sum_i p_i r_i \sin{\theta_i}\cos{\phi_i})-\sum_i p_i f(r_i \sin{\theta_i}\cos{\phi_i}).
\end{equation}
The second derivative of $f(x)$ is
\begin{equation}\label{a16}
\frac{\mathrm{d}^2f(x)}{dx^2}=\frac{e^{-t}}{(1 - e^{-t} x^2)^2 \ln 2}
\end{equation}
Since $0<e^{-t}\leq 1$ and $-1\leq x\leq 1$, we have $\frac{\mathrm{d}^2f(x)}{\mathrm{d}x^2}>0$, so $f(x)$ is a convex function. Since $\sum_i p_i =1$, from Eq. (\ref{a15}), we obtain that $W \leq 0$. Thus, it has been proven that the information gain $I_{a,x}$ monotonically  decreases with the coupling strength $g$.

\end{proof}

\end{document}